\documentclass[10pt,conference]{IEEEtran}
\usepackage{packages}
\begin{document}

\title{\datasetSimple: A Dataset for Opening the Black Box of Cloud Agents on GitHub}

\author{%
\IEEEauthorblockN{%
Jonan Richards\IEEEauthorrefmark{1}, %
Kosei Horikawa\IEEEauthorrefmark{2}, %
Youmei Fan\IEEEauthorrefmark{2}, %
Yutaro Kashiwa\IEEEauthorrefmark{2} and %
Mairieli Wessel\IEEEauthorrefmark{1}%
}
\IEEEauthorblockA{%
\IEEEauthorrefmark{1}Radboud University, the Netherlands, %
\IEEEauthorrefmark{2}Nara Institute of Science and Technology, Japan%
}
}

\maketitle

\begin{abstract}
    Coding agents now complete development tasks in real software projects and submit the results as pull requests. A pull request shows the outcome of an agent's execution but not the process behind it, such as which files the agent read, which commands it ran, or how many tokens it consumed. This is recorded only in the agent's log, and no existing dataset provides such logs of real-world use, including failed agent executions, across a wide range of repositories. We introduce \dataset{}, a large-scale dataset of activity on GitHub's cloud agent platform on public repositories, collected without requiring opt-in from developers. \dataset{} comprises 363,624 tasks and 635,886 sessions, each session an execution of the agent, in 40,565 popular public repositories, with 75,631,136 log entries that record each session step by step, including prompts, intermediate reasoning, tool calls (e.g., file edits, git operations, and GitHub interactions), and token usage, together with the linked pull requests, timelines, workflow runs, and users. Sessions are included regardless of their outcome. \dataset{} enables research on agent behavior and failure modes, efficiency and cost, task formulation, human-agent collaboration, and repository adoption.
    A demonstration video can be found at: \url{https://youtu.be/ZnFpjuK9Azg}
\end{abstract}


\section{Introduction}
Coding agents are now deployed on real software projects, where they complete development tasks and submit the results as pull requests~\cite{li2026AIDevStudyingAI}. A developer assigns an issue or writes a short request, and the agent proceeds autonomously until it opens or updates a pull request. Each session, one execution of the agent, produces two records. The pull request holds the result: a diff, a description, and any discussion that follows. It does not show which files the agent read before editing, which commands it executed, whether it ran the tests and how it responded when they failed, or how many tokens it consumed before completing or abandoning the task. All of this is recorded only in the agent's log. Questions such as why a session resulted in a closed pull request, or what a merged one cost, must therefore be answered from the log rather than from the pull request~\cite{bouzenia2025UnderstandingSoftwareEngineering}.

Existing datasets do not provide logs of agents in real-world use. AIDev collects the pull requests that agents open on public GitHub repositories, together with reviews and comments, but not the logs that produced them~\cite{li2026AIDevStudyingAI}. TraceLab collects logs of developers' local coding agents but removes prompts, tool inputs, and agent output before release~\cite{zhu2026TraceLabCharacterizingCoding}. Logs of agent executions on benchmarks~\cite{pan2025TrainingSoftwareEngineering,badertdinov2026SWErebenchAutomatedPipeline,yang2026SWEsmithScalingData,jain2025R2EGymProceduralEnvironments} are complete, but they are recorded on repositories and tasks chosen and executed by the benchmark authors rather than on tasks that developers assigned to an agent. SWE-chat~\cite{baumann2026SWEchatCodingAgent} is closest to our work, as it records full logs from developers' coding agent use, but it requires developers to install a logging tool and commit its checkpoints, which limits coverage to a few hundred repositories and biases the sample toward successful outcomes.

\dataset{} provides such logs at scale. A developer gives a cloud agent on GitHub's cloud agent platform a task by assigning an issue, mentioning the agent in a pull request comment, or writing a prompt. The platform runs the Copilot cloud agent and third-party agents such as Claude and Codex, although nearly all collected sessions are the Copilot cloud agent (Section~\ref{sec:agents-models}). Each session is one execution of the agent on that task, in a fresh sandbox (Section~\ref{sec:cloud-agent}). On public repositories, the platform exposes every task, session, and log on the Agents tab, which allowed us to collect them without requiring any opt-in from developers. The dataset comprises 363,624 tasks and 635,886 sessions on 40,565 popular repositories, with 75,631,136 log entries. A session log contains the prompt, the agent's reasoning text, every tool call with its arguments and result, and the token usage of each step. The 28,487,431 tool calls cover file views and edits, shell commands, git operations, and comments posted on GitHub. Sessions are included regardless of outcome, including those that ended in an error or whose pull request was never merged. Each session is linked to its pull request and the pull request's timeline events, to the GitHub Actions workflow run that executed it, and to the users involved (Table~\ref{tab:dataset}).

With \dataset{}, a researcher can inspect any individual session to see the developer's request, the agent's actions step by step, and the resulting outcome, or aggregate over hundreds of thousands of sessions. This supports studies of agent behavior and failure modes, of efficiency and cost from per-step token usage, of how developers formulate tasks and respond to results, and of which repositories adopt GitHub's cloud agent platform. We expect the dataset to serve empirical software engineering researchers who study a development process in which agents now participate, as well as researchers and builders of coding agents who seek to understand how agents are used, how they operate, and where they fail.

\section{GitHub's cloud agent platform}
\label{sec:cloud-agent}

GitHub's cloud agent platform runs agents asynchronously in an ephemeral development environment, where they search and edit files, run shell commands such as tests and git, and use GitHub, for example, to open a pull request or reply to a comment.\footnote{\url{https://docs.github.com/en/copilot/responsible-use/agents}} Developers can run the Copilot cloud agent or a third-party agent such as Claude or Codex (Section~\ref{sec:agents-models}).

\begin{figure*}[!t]
    \centering
    \hfill\subfloat[Agents tab\label{fig:session-list}]{%
        \includegraphics[width=0.25\textwidth,trim=0 450 0 0,clip]{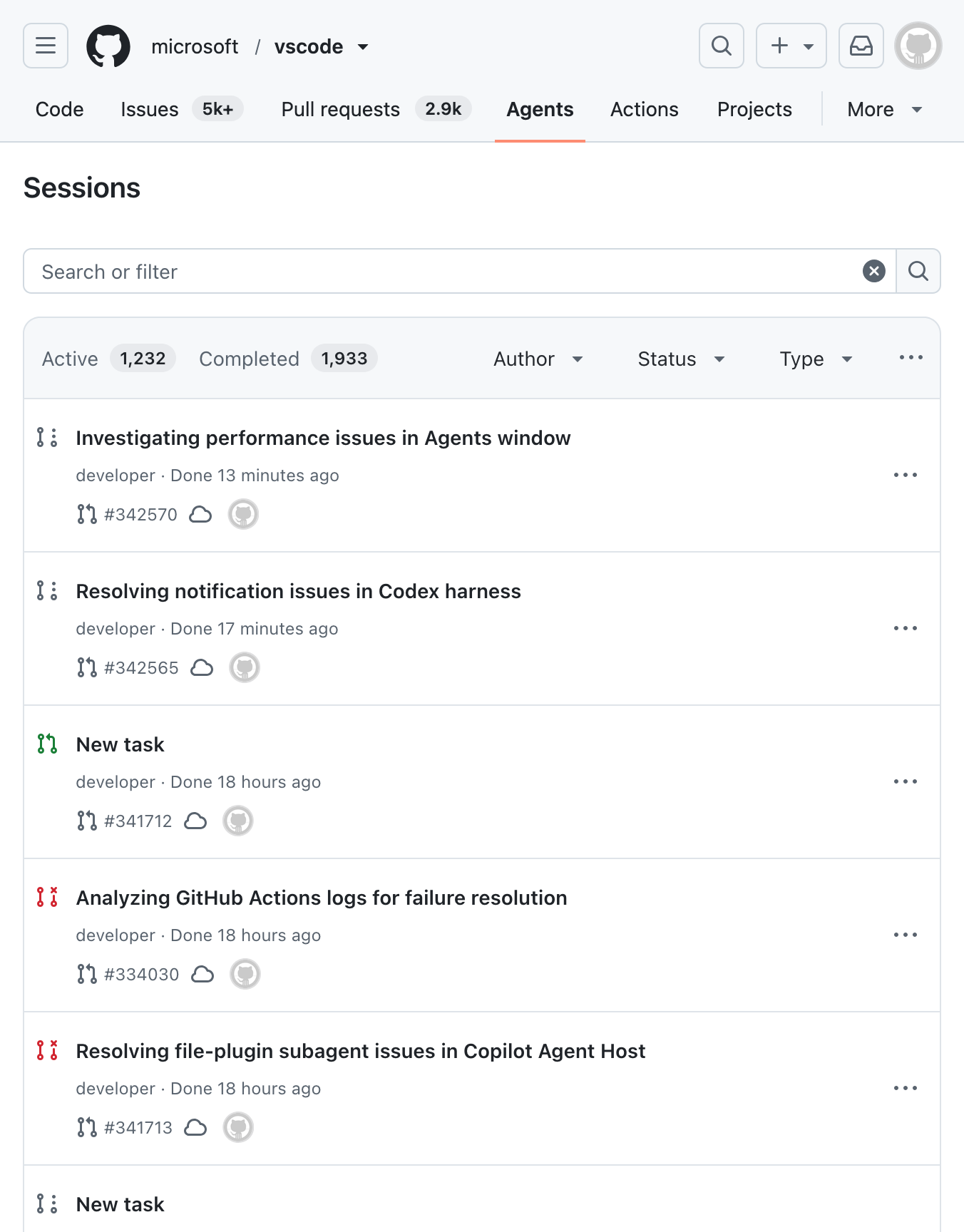}%
    }\hfill
    \subfloat[Task with two sessions\label{fig:session-task}]{%
        \includegraphics[width=0.25\textwidth,trim=0 115 0 135,clip]{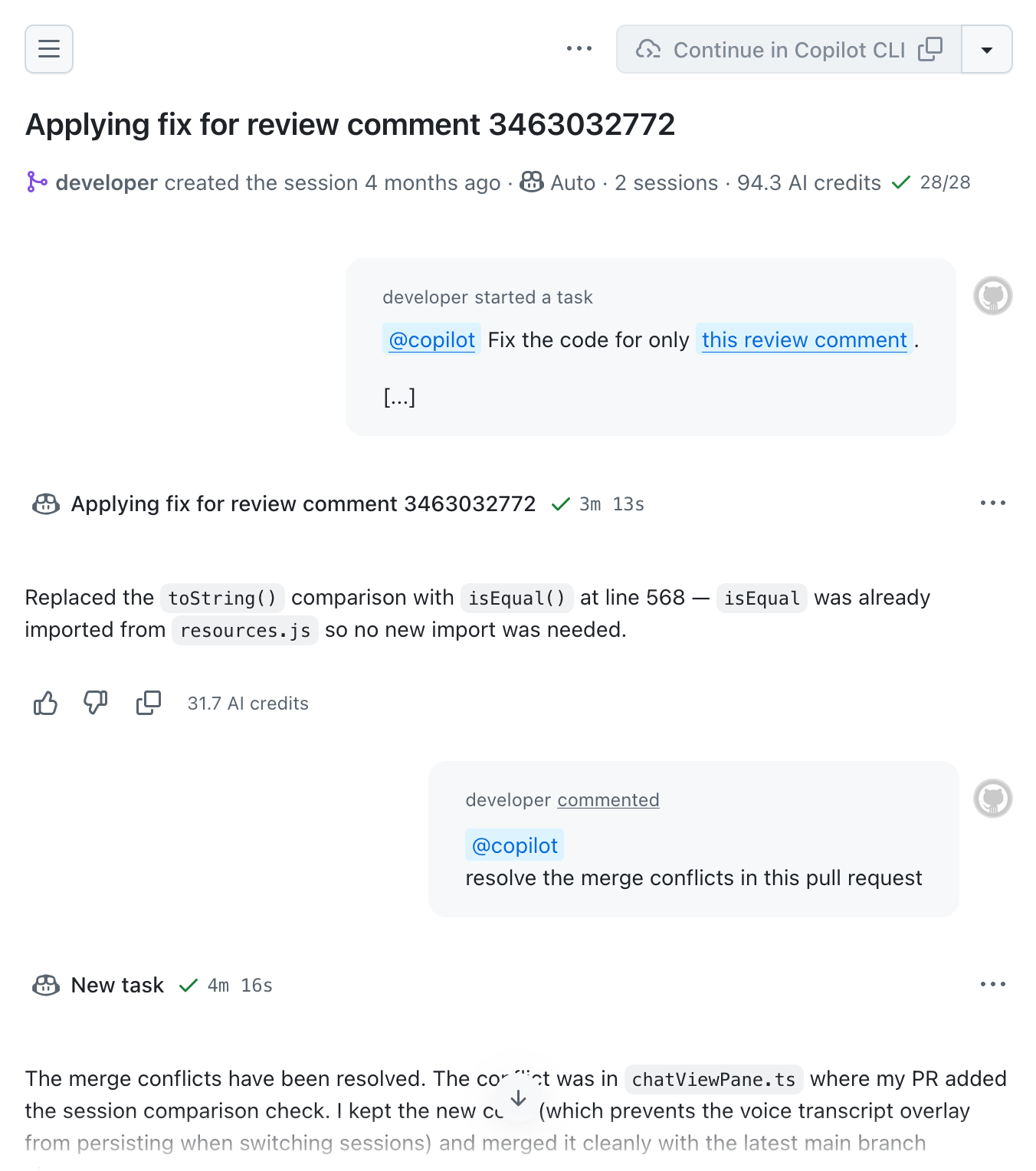}%
    }\hfill
    \subfloat[Session log\label{fig:session-log}]{%
        \includegraphics[width=0.25\textwidth,trim=0 15 0 235,clip]{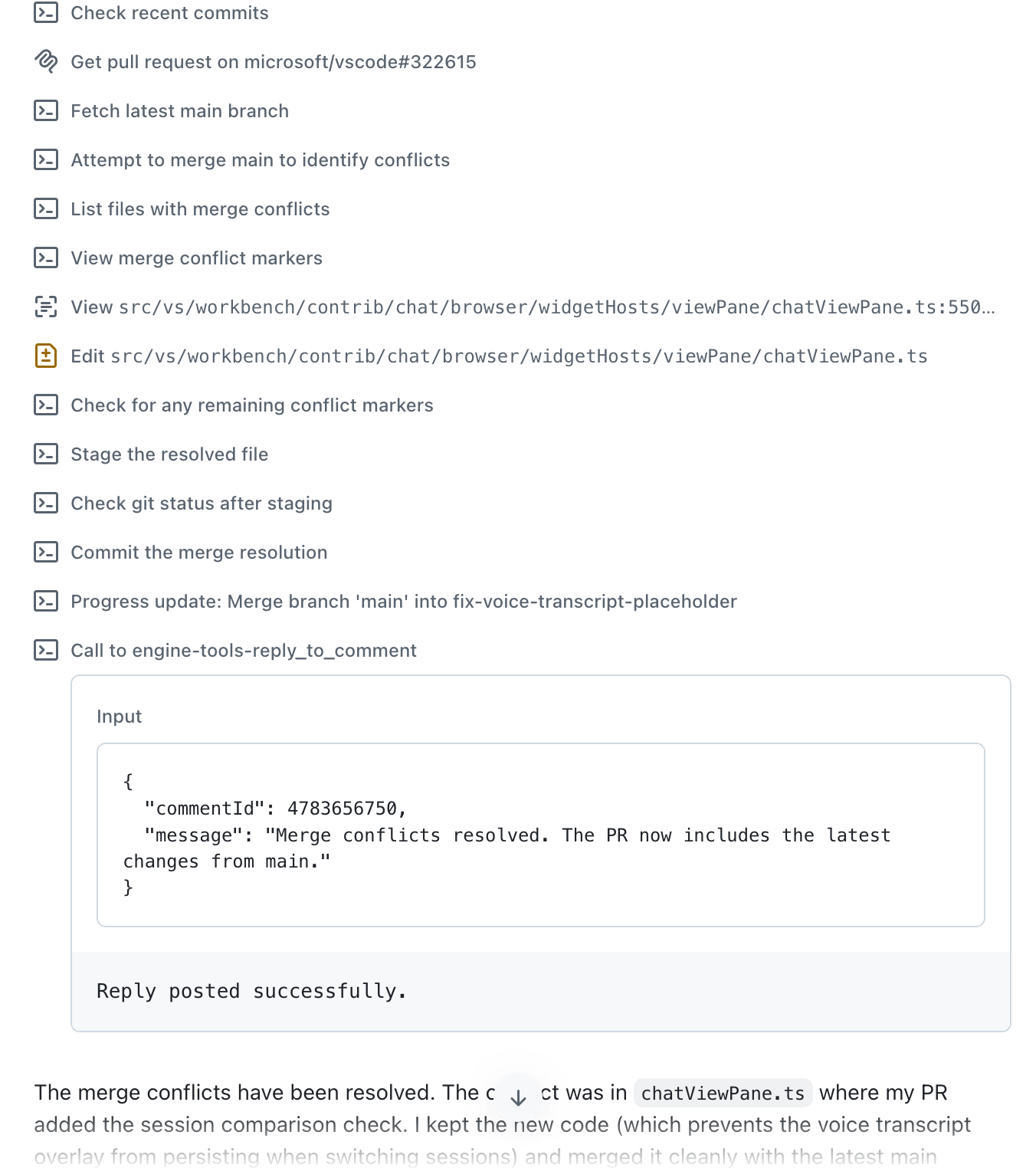}%
    }\hfill{}
    \caption{A Copilot cloud agent task on \texttt{microsoft/vscode}. Usernames and avatars are redacted. (a)~The Agents tab lists recent tasks. (b)~A task with two sessions: a review-comment fix, then a request to resolve the resulting merge conflicts. (c)~Part of the log of the second session.}
  \label{fig:session}
\end{figure*}

Developers can assign a cloud agent a \textit{task}, which typically works on a single branch and a single pull request. That assignment initiates its first \textit{session}, and later requests from the developer add further sessions on the same task (Figure~\ref{fig:session-task}), where each session is one execution of the agent in a fresh sandbox environment. These sessions can be initiated in several ways (Section~\ref{sec:initiation}), e.g., from the Agents tab (Figure~\ref{fig:session-list}), by mentioning the agent, for example \texttt{@copilot}, on a pull request, or from an integration such as Slack or Jira. The \textit{log} of a session is the transcript of its execution, shown in an abridged form in the GitHub web interface (Figure~\ref{fig:session-log}). \dataset{} stores each log entry, including prompts, reasoning traces, tool calls and arguments (Section~\ref{sec:tool-calls}), and token usage.

\section{Related work}
AIDev collects pull requests that local and remote agents such as Codex, Copilot, and Claude Code opened on public GitHub repositories, with their comments and reviews~\cite{li2026AIDevStudyingAI}. AIDev has been used to study agents' real-world outcomes~\cite{ehsani2026WhereAICoding} and how developers interact with agents~\cite{rooijendijk2026WhoSaidCVE}, but cannot show how the agent worked to go from prompt to outcome. DevGPT links ChatGPT conversations that developers shared on GitHub and Hacker News to the artifacts they discuss~\cite{xiao2024DevGPTStudyingDeveloperChatGPT}, but these are chats rather than agent executions with tool calls.

Benchmark datasets record that process as logs of agent executions on fixed tasks. Some take tasks from existing GitHub issues~\cite{pan2025TrainingSoftwareEngineering,badertdinov2026SWErebenchAutomatedPipeline}, others synthesize them~\cite{yang2026SWEsmithScalingData,jain2025R2EGymProceduralEnvironments}. Such logs have been used to study how an agent works through a known task~\cite{bouzenia2025UnderstandingSoftwareEngineering}, but do not reflect how developers use agents, because the benchmark authors chose the repository and task, and started the agent themselves.

In other studies of developers using coding agents, a conversation is often called a session, analogous to a task in \dataset{}. A turn or prompt in that conversation is what we refer to as a session. TraceLab released 4,265 conversations with about 39,200 turns from 43 developers, but removed prompts, tool input, and agent output~\cite{zhu2026TraceLabCharacterizingCoding}. Microsoft researchers characterized logs of 13.5 million conversations with 95.1 million turns in production, but reported only aggregates, without prompts, tool input, or agent output, and did not release their logs~\cite{liu2026AgenticCodingWild}. Closest to \dataset{} is SWE-chat, which includes logs for nearly 18,000 conversations with more than 229,000 user prompts in 707 repositories, together with line-level authorship of the committed code~\cite{baumann2026SWEchatCodingAgent}. It only covers developers who installed Entire, a local logging tool, and committed its checkpoints.

\dataset{} records public activity on GitHub's cloud agent platform without additional opt-in or a committed checkpoint, reducing bias toward successful sessions. Compared to SWE-chat, it collected agent logs from $50.4\times$ more repositories, with $20.2\times$ more tasks than SWE-chat's sessions and $2.8\times$ more sessions than its user prompts.

\section{The \dataset{} dataset}
\subsection{Data collection}
We collected sessions on GitHub's cloud agent platform on public repositories. These sessions execute on GitHub's infrastructure, and each task, prompt, and log is already visible on the repository's Agents tab. Locally run sessions and private repositories are not collected.

The seed is SEART GitHub Search Engine (GHS)~\cite{dabic2021SamplingProjectsGithub}, a dataset of 1,997,424 public repositories with at least 10 stars, accessed on September 28, 2026. For each repository we listed agent tasks, including archived tasks, with the GitHub REST API. The documented REST task omits fields such as session trigger and workflow run, so tasks and sessions were fetched from an undocumented endpoint of the GitHub Copilot API.\footnote{\url{https://api.githubcopilot.com}} The GitHub Acceptable Use Policies explicitly allow any method of obtaining publicly available data.\footnote{\url{https://docs.github.com/tos}} Session logs come from the same Copilot API as a server-sent event stream. More than 99.99\% of events parsed as JSON; the rest were kept as raw strings. We also collected the sessions' workflow runs, the pull requests of tasks and sessions with their timeline events, and all linked users. As data minimization, user records kept only the login, GitHub id, and account creation date, not profile fields such as name, email, or bio.

Repositories no longer available through the REST task endpoint, e.g., after deletion, renaming, or being made private, are left out. Repositories with no tasks are kept. Tasks, pull requests, workflow runs, timeline events, and users that the API does not return are left out as well, although other tables may still refer to them. A session whose log is not found is kept, with no log entries.

\subsection{Internal structure}
\begin{table*}[!tb]
    \centering
    \caption{Summary of the \dataset{} dataset}
    \label{tab:dataset}
    \begin{tabularx}{\textwidth}{@{}l r r l >{\raggedright\arraybackslash}X@{}}
\toprule
&
\textbf{\# Records} &
\textbf{Size} &
\textbf{Table} &
\textbf{Content} \\
\midrule

\textbf{Repositories} &
1,861,300 &
462.7 MB &
\verb/repositories/ & \multirow[t]{2}{=}{Public GitHub repositories with at least 10 stars (incl. name, size, language, stars, forks, labels, topics).} \\
\hspace{1em}-- 2.18\% with agent tasks & 40,565 & & & \\
\addlinespace[1.6pt]

\textbf{Agent tasks} &
363,624 &
83.7 MB &
\verb/agent_tasks/ & Agent assignments on a repository (incl. creator, name, state, sessions, branches, pull requests). \\
% \hspace{1em}-- 97.20\% with sessions & 353,432 & & & \\
\addlinespace[1.6pt]

\textbf{Agent sessions} &
635,886 &
243.1 MB &
\verb/agent_sessions/ & Executions of an agent on a task (incl. user, triggering event, model, prompt, usage, workflow run, pull request). \\
% \hspace{1em}-- 98.24\% with log & 624,668 & & & \\
% \hspace{1em}-- 90.09\% with pull request & 572,840 & & & \\
% \hspace{1em}-- 87.20\% with workflow run & 554,496 & & & \\
\addlinespace[1.6pt]

\textbf{Log entries} &
75,631,136 &
63.0 GB &
\verb/agent_session_logs/ & Session log events: chat-completion chunks (incl. reasoning, tool calls, token usage) and user and tool messages. \\
\addlinespace[1.6pt]

\textbf{Pull requests} &
317,878 &
483.9 MB &
\verb/pull_requests/ & Pull requests worked on in tasks and sessions (incl. title, author, state, branches, change size). \\
% \hspace{1em}-- 99.82\% with timeline events & 317,297 & & & \\
\addlinespace[1.6pt]

\textbf{Pull request timeline events} &
9,070,939 &
327.4 MB &
\verb/pull_request_timelines/ & Timeline events for a pull request (incl. event type, actor, label, commit SHA, message). \\
\addlinespace[1.6pt]

\textbf{Workflow runs} &
554,496 &
83.3 MB &
\verb/workflow_runs/ & GitHub Actions workflow runs that executed an agent session (incl. status, conclusion, triggering user). \\
\addlinespace[1.6pt]

\textbf{Users} &
857,428 &
211.5 MB &
\verb/users/ & Users, organizations, and bots related to the other tables (incl. GitHub ID, login slug, creation date). \\
% \hspace{1em}-- 4.40\% with tasks or sessions & 37,714 & & & \\

\midrule
\textbf{Total} & \textbf{89,292,687} & \textbf{64.9 GB} & & \\
\bottomrule
\end{tabularx}

\end{table*}

\dataset{} is a set of Parquet tables (Table~\ref{tab:dataset}), sharded into chunks of $\sim$80--120MB so each table can be downloaded and scanned in parallel. Records store the IDs of related rows.

Repository rows are the SEART GHS seed data, with links to their tasks. Tasks and sessions largely match the documented REST task resource. Pull requests, workflow runs, timeline events, and users follow their GitHub REST schema. Log entries are a tool result, user message, or an OpenAI-format chat-completion chunk with assistant text, reasoning, tool calls (Section~\ref{sec:tool-calls}), and token usage, plus fields such as billing, warnings, and tool telemetry. The schema documentation (Section~\ref{sec:access}) maps each column to its source field.

\subsection{How to access}
\label{sec:access}

The \dataset{} dataset is available on Hugging Face.\footnote{\url{https://huggingface.co/datasets/risenlab/agentlogs}} The data collection pipeline, schema documentation, a data sample (4 tasks, 4 sessions, 458 log entries), and example notebooks that query the dataset with DuckDB, Polars, or a low-memory typed streaming reader are available on GitHub.\footnote{\url{https://github.com/risenlab/agentlogs}} Python type definitions for every table are available as the PyPI package \texttt{risenlab-agentlogs}. Code is licensed under MIT and the dataset under CC~BY~4.0. The released pipeline allows re-collection, and we plan to publish updated snapshots on a best-effort basis.

\section{Preliminary analysis}
% To characterize the dataset, we describe how sessions start and how many have been created over time, and which tools, agents, and models GitHub's cloud agent platform runs.

\subsection{Initiating a session}
\label{sec:initiation}

Almost all sessions (99.9\%) record their trigger in the \texttt{event} field. Sessions can be started on a pull request, for example by mentioning \texttt{@copilot} in a comment (29.5\% of sessions) or a review (12.8\%), by requesting a Copilot code review (5.6\%), or by requesting fixes for review feedback (1.6\%). Developers can also enter a prompt, for example in the GitHub web interface (15.9\%), the mobile app (2.5\%), an IDE (0.6\%), or the CLI (0.2\%). Developers can assign an issue (18.0\%), including on a tracker outside GitHub. Some sessions are initiated through the GitHub API, Copilot chat, or an integration such as Slack, Teams, Sentry, or an MCP tool (10.8\%), others are triggered by a failed GitHub Actions workflow run (1.9\%), or by a security, quality, or coverage alert (0.3\%). The remaining sessions (0.1\%) have a rare trigger, such as a scheduled workflow run.

\subsection{Sessions over time}
\begin{figure}[!t]
    \centering
    \includegraphics[width=0.85\columnwidth]{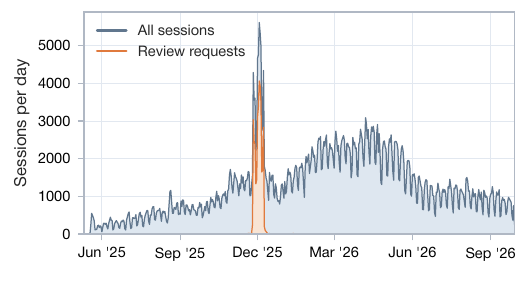}
    \caption{Agent sessions per day. Blue shows all sessions and orange shows the share of sessions with a \texttt{review-request} event type.}
    \label{fig:sessions-per-day}
\end{figure}

Figure~\ref{fig:sessions-per-day} shows sessions created per day, from the earliest session on May 11, 2025, to the cutoff on September 29, 2026. The spike is \texttt{review-request} sessions: from November 24 through December 8, 35,080 of 54,299 sessions have that event type, and the type is rare outside that period. Copilot code review remains a feature on GitHub, suggesting that GitHub only briefly recorded review requests as cloud agent sessions. Disregarding this spike, sessions per day rose to a peak of about 3,000 in early April 2026 and then declined to about 840 per day by September.

\subsection{Tool calls}
\label{sec:tool-calls}
One log entry can contain several tool calls. \dataset{} contains 28,487,431 tool calls, for an average of 45.6 calls per session with a non-empty log. File operations such as \texttt{view}, \texttt{edit}, and \texttt{grep} make up most of the tool calls (39.9\%), together with shell interaction (e.g. \texttt{bash} and \texttt{read\_bash}; 37.4\%). Also common are tool calls related to setup and progress (14.2\%), operations on GitHub such as commenting and reading issues (3.8\%), and code review and scanning tools (2.1\%). The rest include subagent delegation, planning, memory, and browser interaction.

\subsection{Agents and models}
\label{sec:agents-models}
The workflow-run user, not the task, records which agent ran. Of 554,496 workflow runs, 98.8\% are \texttt{copilot-swe-agent}, 0.8\% \texttt{anthropic-code-agent}, and 0.2\% \texttt{openai-code-agent}; every other account, including third-party agents such as Bright Security, is at most 0.1\%.

Each session records the model that ran. Claude accounts for 56.4\% of sessions and GPT for 17.5\%, with the most common versions being \texttt{claude-sonnet-4.6} (25.9\% of sessions), \texttt{claude-sonnet-4.5} (17.4\%), and \texttt{gpt-5.3-codex} (9.5\%). MAI, Gemini, Kimi, Grok, and other models each account for 0.1\% of sessions or less. The model is unrecorded for 25.3\% of sessions.

\section{Discussion}
The preliminary analysis shows the scale and variety of the activity in \dataset{} from May 2025 through September 2026, with sessions triggered by pull requests, issues, prompts, integrations, and other events, and 28,487,431 tool calls spanning file operations, shell commands, GitHub interactions, code review, and subagents.

However, \dataset{} has several limitations. Although third-party agents can run on GitHub's cloud agent platform, nearly all sessions are the Copilot cloud agent, so comparisons across sessions are concentrated on one agent. The model is unrecorded for 25.3\% of sessions, reducing the number of sessions useful for comparisons that split by model.

Session context, such as a custom Markdown instruction file, the triggering text, or a pull request comment thread, is often not stored on the session record. Commits are not collected, and timeline events are not tied to sessions or log entries. Session context, commits, and timeline events often appear in the log instead, as a prompt to the agent or as a tool call, but may be hard to extract reliably.

Still, no other released dataset includes the prompts, reasoning, and tool calls of developer-initiated sessions, including sessions that fail, across public repositories at this scale. \dataset{} uniquely enables researchers to answer questions about agent behavior, such as which patterns agents follow and how those patterns differ between successful and failed sessions, about collaboration in the GitHub ecosystem, such as how developers initiate sessions and how triggers and repository context relate to the agent's execution and outputs, and about repository adoption, such as what characterizes the 2.18\% of popular repositories that adopt the platform, compared to those that do not.

\bibliographystyle{IEEEtran}
\bibliography{references}

@inproceedings{dabic2021SamplingProjectsGithub,
  title = {Sampling Projects in {Github} for {{MSR}} Studies},
  booktitle = {2021 {{IEEE}}/{{ACM}} 18th {{International Conference}} on {{Mining Software Repositories}} ({{MSR}})},
  author = {Dabic, Ozren and Aghajani, Emad and Bavota, Gabriele},
  year = 2021,
  pages = {560--564},
  publisher = {IEEE},
  urldate = {2026-07-17}
}

@article{badertdinov2026SWErebenchAutomatedPipeline,
  title = {{{SWE}}-Rebench: {{An Automated Pipeline}} for {{Task Collection}} and {{Decontaminated Evaluation}} of {{Software Engineering Agents}}},
  shorttitle = {{{SWE}}-Rebench},
  author = {Badertdinov, Ibragim and Golubev, Alexander and Nekrashevich, Maksim and Shevtsov, Anton and Karasik, Simon and Andriushchenko, Andrei and Trofimova, Maria and Litvintseva, Daria and Yangel, Boris},
  year = 2026,
  month = apr,
  journal = {Advances in Neural Information Processing Systems},
  volume = {38},
  doi = {10.52202/085713-0788},
  urldate = {2026-10-21},
  langid = {english}
}

@misc{jain2025R2EGymProceduralEnvironments,
  title = {{{R2E-Gym}}: {{Procedural Environments}} and {{Hybrid Verifiers}} for {{Scaling Open-Weights SWE Agents}}},
  shorttitle = {{{R2E-Gym}}},
  author = {Jain, Naman and Singh, Jaskirat and Shetty, Manish and Zheng, Liang and Sen, Koushik and Stoica, Ion},
  year = 2025,
  month = apr,
  number = {arXiv:2504.07164},
  eprint = {2504.07164},
  primaryclass = {cs.SE},
  publisher = {arXiv},
  doi = {10.48550/arXiv.2504.07164},
  urldate = {2026-10-21},
  archiveprefix = {arXiv},
  howpublished = {arXiv:2504.07164}
}

@misc{pan2025TrainingSoftwareEngineering,
  title = {Training {{Software Engineering Agents}} and {{Verifiers}} with {{SWE-Gym}}},
  author = {Pan, Jiayi and Wang, Xingyao and Neubig, Graham and Jaitly, Navdeep and Ji, Heng and Suhr, Alane and Zhang, Yizhe},
  year = 2025,
  month = jun,
  number = {arXiv:2412.21139},
  eprint = {2412.21139},
  primaryclass = {cs.SE},
  publisher = {arXiv},
  doi = {10.48550/arXiv.2412.21139},
  urldate = {2026-10-21},
  archiveprefix = {arXiv},
  howpublished = {arXiv:2412.21139}
}

@article{yang2026SWEsmithScalingData,
  title = {{{SWE}}-Smith: {{Scaling Data}} for {{Software Engineering Agents}}},
  shorttitle = {{{SWE}}-Smith},
  author = {Yang, John and Lieret, Kilian and Jimenez, Carlos and Wettig, Alexander and Khandpur, Kabir and Zhang, Yanzhe and Hui, Binyuan and Press, Ofir and Schmidt, Ludwig and Yang, Diyi},
  year = 2026,
  month = apr,
  journal = {Advances in Neural Information Processing Systems},
  volume = {38},
  doi = {10.52202/085713-3239},
  urldate = {2026-10-21},
  langid = {english}
}

@inproceedings{bouzenia2025UnderstandingSoftwareEngineering,
  title = {Understanding {{Software Engineering Agents}}: {{A Study}} of {{Thought-Action-Result Trajectories}}},
  shorttitle = {Understanding {{Software Engineering Agents}}},
  booktitle = {2025 40th {{IEEE}}/{{ACM International Conference}} on {{Automated Software Engineering}} ({{ASE}})},
  author = {Bouzenia, Islem and Pradel, Michael},
  year = 2025,
  month = nov,
  pages = {2846--2857},
  issn = {2643-1572},
  doi = {10.1109/ASE63991.2025.00234},
  urldate = {2026-10-21}
}

@inproceedings{li2026AIDevStudyingAI,
  title = {{{AIDev}}: {{Studying AI Coding Agents}} on {{GitHub}}},
  shorttitle = {{{AIDev}}},
  booktitle = {Proceedings of the 23rd {{International Conference}} on {{Mining Software Repositories}}},
  author = {Li, Hao and Zhang, Haoxiang and Hassan, Ahmed E.},
  year = 2026,
  month = jul,
  series = {{{MSR}} '26},
  pages = {1029--1033},
  publisher = {Association for Computing Machinery},
  address = {New York, NY, USA},
  doi = {10.1145/3793302.3797249},
  urldate = {2026-10-20},
  isbn = {979-8-4007-2474-9}
}

@incollection{ehsani2026WhereAICoding,
  title = {Where {{Do AI Coding Agents Fail}}? {{An Empirical Study}} of {{Failed Agentic Pull Requests}} in {{GitHub}}},
  shorttitle = {Where {{Do AI Coding Agents Fail}}?},
  booktitle = {Proceedings of the 23rd {{International Conference}} on {{Mining Software Repositories}}},
  author = {Ehsani, Ramtin and Pathak, Sakshi and Rawal, Shriya and Al Mujahid, Abdullah and Imran, Mia Mohammad and Chatterjee, Preetha},
  year = 2026,
  month = jul,
  pages = {807--811},
  publisher = {Association for Computing Machinery},
  address = {New York, NY, USA},
  urldate = {2026-10-20},
  isbn = {979-8-4007-2474-9}
}

@inproceedings{rooijendijk2026WhoSaidCVE,
  title = {Who {{Said CVE}}? {{How Vulnerability Identifiers Are Mentioned}} by {{Humans}}, {{Bots}}, and {{Agents}} in {{Pull Requests}}},
  shorttitle = {Who {{Said CVE}}?},
  booktitle = {Proceedings of the 23rd {{International Conference}} on {{Mining Software Repositories}}},
  author = {Rooijendijk, Pien and Treude, Christoph and Wessel, Mairieli},
  year = 2026,
  month = jul,
  series = {{{MSR}} '26},
  pages = {994--998},
  publisher = {Association for Computing Machinery},
  address = {New York, NY, USA},
  doi = {10.1145/3793302.3793616},
  urldate = {2026-10-20},
  isbn = {979-8-4007-2474-9}
}

@misc{baumann2026SWEchatCodingAgent,
  title = {{{SWE}}-Chat: {{Coding Agent Interactions From Real Users}} in the {{Wild}}},
  shorttitle = {{{SWE}}-Chat},
  author = {Baumann, Joachim and Padmakumar, Vishakh and Li, Xiang and Yang, John and Yang, Diyi and Koyejo, Sanmi},
  year = 2026,
  month = oct,
  publisher = {arXiv},
  doi = {10.48550/ARXIV.2604.20779},
  urldate = {2026-10-21},
  copyright = {Creative Commons Attribution 4.0 International},
  langid = {english}
}

@misc{liu2026AgenticCodingWild,
  title = {Agentic {{Coding}} in the {{Wild}}: {{Characterizing GitHub Copilot Traces}} at {{Production Scale}}},
  shorttitle = {Agentic {{Coding}} in the {{Wild}}},
  author = {Liu, Banruo and Qiu, Haoran and Goiri, {\'I}{\~n}igo and Fonseca, Rodrigo and Bianchini, Ricardo and Choukse, Esha},
  year = 2026,
  month = jul,
  journal = {arXiv.org},
  urldate = {2026-10-21},
  howpublished = {https://arxiv.org/abs/2608.00101v1},
  langid = {english}
}

@misc{zhu2026TraceLabCharacterizingCoding,
  title = {{{TraceLab}}: {{Characterizing Coding Agent Workloads}} for {{LLM Serving}}},
  shorttitle = {{{TraceLab}}},
  author = {Zhu, Kan and Jacob, Mathew and Ma, Chenxi and Pan, Yi and Wang, Stephanie and Krishnamurthy, Arvind and Kasikci, Baris},
  year = 2026,
  month = jun,
  journal = {arXiv.org},
  urldate = {2026-10-21},
  howpublished = {https://arxiv.org/abs/2606.30560v2},
  langid = {english}
}

@inproceedings{xiao2024DevGPTStudyingDeveloperChatGPT,
  title = {{{DevGPT}}: {{Studying}} Developer-{{ChatGPT}} Conversations},
  booktitle = {Proceedings of the 21st International Conference on Mining Software Repositories},
  author = {Xiao, Tao and Treude, Christoph and Hata, Hideaki and Matsumoto, Kenichi},
  year = 2024,
  series = {{{MSR}} '24},
  pages = {227--230},
  publisher = {Association for Computing Machinery},
  address = {New York, NY, USA},
  doi = {10.1145/3643991.3648400},
  isbn = {979-8-4007-0587-8}
}

\end{document}